\documentclass[twocolumn,showpacs,preprintnumbers,amsmath,amssymb]{revtex4}

\usepackage{graphicx}
\usepackage{dcolumn}
\usepackage{bm}

\usepackage{epstopdf}

\begin{document}


\title[Absence of reflectivity of the phonon-polariton at the SiC surface from the metal mask edge]{Absence of reflectivity of the phonon-polariton at the SiC surface from the metal mask edge}
\author{V.S. Ivchenko}
\author{D.V. Kazantsev}%
 \altaffiliation[Also at ]{HSE University,
Moscow, Russia 101000.}
 \email{kaza@itep.ru}
\author{V.A. Ievleva}
\affiliation{P.N. Lebedev Physical Institute of the Russian Academy of Sciences, 119991, Moscow, Russia
}%
\author{D.A. Matienko}
\affiliation{P.N. Lebedev Physical Institute of the Russian Academy of Sciences, 119991, Moscow, Russia
}%
\author{E.A. Kazantseva}
\affiliation{Moscow Technological University,
Moscow, Russia 119454
}%
\author{A.Yu. Kuntsevich}

\affiliation{P.N. Lebedev Physical Institute of the Russian Academy of Sciences, 119991, Moscow, Russia
}%

\date{\today}

\begin{abstract}
Surface phonon polariton (SPhP) waves are excited at the silicon
carbide ($SiC$) surface under irradiation of light close to the
lattice resonance frequency. Metal mask at the surface blocks
irradiation of certain areas and thus allows tuning standing or
propagating wave pattern and, thus, open opportunities for surface
polariton optic devices. In this study we show by means of scanning
near-field microscopy that the edge of such a mask reflects SPhP
waves negligibly. This condition differs dramatically from numerous
recent observations of polariton reflections in 2D materials and
makes the metallized $SiC$ platform advantageous in a sense of
capability to calculate wavefield using a simple Green
function-based approach.
\end{abstract}

\pacs{07.79.Fc, 68.37.Ps, 07.60.-j, 87.64.Je, 61.46.+w, 85.30.De, 68.65.Pq}
\maketitle

{\bf Introduction.} Surface phonon polariton (SPhP) waves could be
excited by external irradiation at the surfaces of polar crystals
and are therefore of great interest for on-chip surface optic
applications~\cite{Basov_PolaritonPanorama_NanoPhot_2021}. SPhPs
emerge if the frequency is close to the optical phonon frequency
$\omega_T$, i.e., mechanical resonance of the lattice, where local
dielecrtic permittivity reaches several hundreds
~\cite{Sasaki_SiC_Raman_PRB1989,Eps_SiC_Landolt_B}. A convenient and
commonly accepted description of near-surface electromagnetic wave
propagation is a simultaneous solution of the Maxwell equations for
electromagnetic fields and the Newton equations for lattice atoms
\cite{LST_1941,Barron_PR1961,Rup_Englm_RPP1970,Mills_Burst_RPP1974}.

\begin{figure}
\resizebox{0.45\textwidth}{!}
{\includegraphics{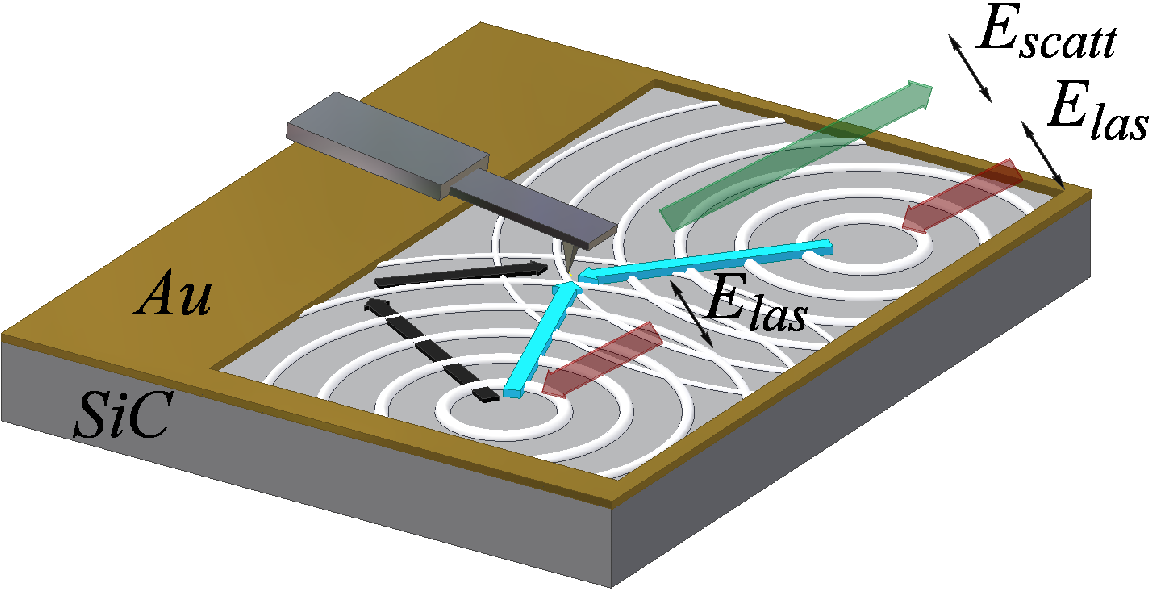}} \caption{Excitation of
phonon-polariton waves on an open $SiC$ surface by external
irradiation. A phenomenon model assumes the use of the Green's
function. Laser (transparent red arrows) light with constant
amplitude $E_{las}$ affects all points of the $SiC$ surface not
covered by the $Au$ mask and generates concentrically diverging
waves from them. These waves sum up at the point of the tip
location, having arrived at it along the straight trajectories,
indicated by light arrows. Under the action of the local field, the
tip scatters a wave (transparent green arrow) with an amplitude
$E_{scatt}$ back into the instrument. $E_{scatt}$ value is
proportional to the total field under it. SPhP rays propagating from
their point of excitation to the probe with a reflection (shown as
black arrows) would significantly complicate the calculations.}
\label{fig:sSNOM_Sample_n_Tip}
\end{figure}

This model leads to the same equation as for the plasma of itinerant
electrons in metals. Numerically, however, plasmon resonances in
metals lie in the visible or UV spectral range owing to small
electron mass, while SPhPs are located in the middle and far IR
spectral range. For example, in $SiC$ the resonance occurs at
$\nu_T(SiC)=795$ {cm}$^{-1}$. The necessity to use dedicated IR
lasers and photodetectors made surface phonon-polariton phenomena
much less studied than plasmon-polaritonics.

Similarly to surface plasmons, the SPhP waves are located in a thin
layer close to the interface and could propagate in a lateral
direction without much damping. It was previously demonstrated by
scanning near-field optical microscopy (SNOM) methods that phonon
polariton fields are influenced by patterns made of metal (usually
gold) films or dielectric. Indeed, the field of incident wave does
not reach the $SiC$ surface under metal, (see
Fig.\ref{fig:sSNOM_Sample_n_Tip}) while the dielectric film changes
substantially the media interface boundary conditions. Depending on
the design of the mask, various surface wave patterns are formed,
including edge-launched
waves~\cite{Huber_SiC_SPP_dispersion_APL2005, Mancini_AcsPhot_2022},
resonators~\cite{UltraConfined_PhPolarit_NatComm_2020}, single metal
spot scatterers~\cite{Hillenbr_Round_Island_SiC_JM2008}, concave
mirrors~\cite{Hillenbr_Horseshoe_APL2008}, and diffraction
gratings\cite{Kaza_Bragg_APL_2024}.

In recent years a great deal of interest arose to the surface IR
plasmon or phonon-polariton electromagnetic waves in 2D materials,
thin films, and heterostructures
\cite{Hillenbr_NatLett_Graphene_2012,
Basov_Graphene_on_BN_NatNano_2015, SurfPhPolarit_WdW_NatComm_2018,
UltraConfined_PhPolarit_NatComm_2020}. SNOM studies in these papers
reveal experimental observations of numerous phenomena related to
the scattering or reflection of the surface waves at the boundaries
of the mesa-defined elements or flakes. In particular, interference
of the wave at the tip near the boundary with the reflected surface
waves produces typical oscillations - a direct proof of the surface
waves. For the surface of a bulk $SiC$, the tip plays a negligible
role in excitation and propagation of the SPhP electromagnetic
field, except for a rather thin (100-900nm) layer
~\cite{Mancini_AcsPhot_2022}. Nevertheless some observed wave
patterns on metallized $SiC$ surface were explained by the
generation of the surface wave by the metal mask
edge~\cite{Hillenbr_Horseshoe_APL2008}.

Our paper experimentally demonstrates an important yet not evident
fact: the absence of phonon-polariton wave reflection at the edge of
a metallic mask on the $SiC$ surface using SNOM methods. This
feature makes polaritonics at nearly-resonant surfaces of polar
crystals different from that in thin layers and heterostructures. If
the reflection of SPhP waves from the edge took place in $SiC$, then
the calculation of the field configuration in the structures, with
subsequent modeling and prediction of the properties, would be a
very difficult task. The absence of reflection from the metal mask
allows us to simplify the calculations within the semi-analytical
Green-function-based method, that is crucial for modeling SPhP
devices.

{\bf Experimental methods.} The surface electromagnetic wave field
is investigated using a scattering-type apertureless scanning
near-field optical microscope (sSNOM)
~\cite{ASNOM_Wickramasinghe_PatentUSA_1990, s_SNOM_first} (see
fig.\ref{fig:sSNOM_Sample_n_Tip}) under normal conditions.
Metallized $MFG01-Pt$ probes with a mechanical resonance frequency
of $55-60$ kHz were used; the amplitude of oscillations of the tip
height above the sample was 50-70~nm. The signal beam of the
Michelson scheme was focused by an objective onto the probing tip
and onto the $SiC$ sample around it.The focal spot size was about
$50-80 \mu m$ around the tip at a wavelength of $\lambda\sim 10.7
\mu m$. The spatial resolution of sSNOM is determined by the radius
of the probe tip (below 30~nm). The signal extracted by sSNOM during
optical homodyning ~\cite{s_SNOM_first,
Labardi_SecondHarm_ASNOM_APL2000, Keilmann_PTRS2004_sSNOM}, with its
amplitude and phase, is proportional to the complex amplitude of the
local electromagnetic field $E_{loc}(x,y)$ at the point of the tip
location $(x,y)$, as  found for SPhP waves on the surface of $SiC$
~\cite{Huber_SiC_SPP_dispersion_APL2005, Kaza_JETP2006_Engl,
Kaza_Spiral_Trajectory_APA_2013}.

{\bf Theoretical background.} It was previously
suggested~\cite{Kaza_JETP2006_Engl,
Kaza_FieldEnhancement_JETPLe_2018} that in some simple cases, the
running and standing polariton waves excited by incident light on
the $SiC$ surface can be described by integrating the Green's
function, similar to
plasmonics~\cite{Zayats_Plasmonics_PhysRep_2005,
Krenn_PlasmonGF_PRL_1999}. Within this approach, amplitude of the
laser field under the metal mask is zero, whereas uncovered areas
are sources of diverging SPhP waves, with the amplitude and phase
determined by the amplitude and phase of the external laser field
$E_{las}(x',y')$, where $(x',y')$ are coordinates. All such waves
propagate along the surface to the observation point $(x,y)$, where
the probing tip is located. Thus, the probing tip dipole is driven
by a local field $E_{loc}(x,y)=E_{las}(x,y)+E_{SPhP}(x,y)$. The
second term in this expression can be expressed as an integral \
\begin{align}
 E_{SPhP}(x,y)=\xi \int \limits_{SiC} E_{las}(x',y')G(\Delta r_{xy})dx'dy',
 \label{eq:GF_Integral}
 \end{align}
where the index $SiC$ implies the integration over the open $SiC$
surface from which the polariton wave can propagate to the probe tip
along a straight line. The Green's function is denoted as $G()$, and
its argument $\Delta r_{xy}$ means the distance from each elementary
source $(x',y')$ to the tip $(x,y)$. The complex Green's function is
the Hankel function $H_0(k_{xy}(\omega)\Delta r_{xy})$, with an
argument corresponding to the eigenvalue $k_{xy}(\omega)$ found by
solving the homogeneous wave equations for a surface
polariton~\cite{LST_1941,Barron_PR1961, Rup_Englm_RPP1970,
Mills_Burst_RPP1974}:

 \begin{equation}
 k_{xy}(\omega)=\frac{\omega}{c}\sqrt{\frac{\varepsilon_{ab}\varepsilon_{bn}
 (\omega)} {\varepsilon_{ab}+\varepsilon_{bn}(\omega)}}
 \label{eq:k_xy}
\end{equation}\\

The indices $ab$ and $bn$ denote the dielectric constant of the
medium above and below the interface, respectively. Above the
interface, a vacuum is usually assumed ($\varepsilon_{vac}\equiv
1$). Below the interface, a polar crystal is located, the analytical
expression for the dielectric function of
which~\cite{LST_1941,Barron_PR1961, Rup_Englm_RPP1970,
Mills_Burst_RPP1974} can be written as follows:
\begin{equation}
 \varepsilon_{bn} (\omega)=\varepsilon_{\infty}
 \left(
 1+\frac{\omega^{2}_{L}-\omega^{2}_{T}}{\omega^{2}_{T}-\omega^{2}-i\omega\Gamma}
 \right)
 \label{eq:EpsilonSiC}
\end{equation}

The values of $\omega_T$, $\omega_L$, and $\Gamma$ can be determined
independently from the Raman spectra; see,
e.g.,~\cite{RamanSiC_PR_1968,Sasaki_SiC_Raman_PRB1989,Harima_AP-1995_Raman}.
The parameter $\varepsilon_{\infty}$ is the dielectric function at a
frequency well above the lattice vibration frequencies. The function
$H_0(k_{xy}(\omega)\Delta r_{xy})$ is a solution of the wave
equation~\cite{LST_1941,Barron_PR1961, Rup_Englm_RPP1970,
Mills_Burst_RPP1974} in cylindrical coordinates, so the integral
(\ref{eq:GF_Integral}) must also be a solution of the wave equation.

\begin{figure}
\resizebox{0.45\textwidth}{!}
 {\includegraphics{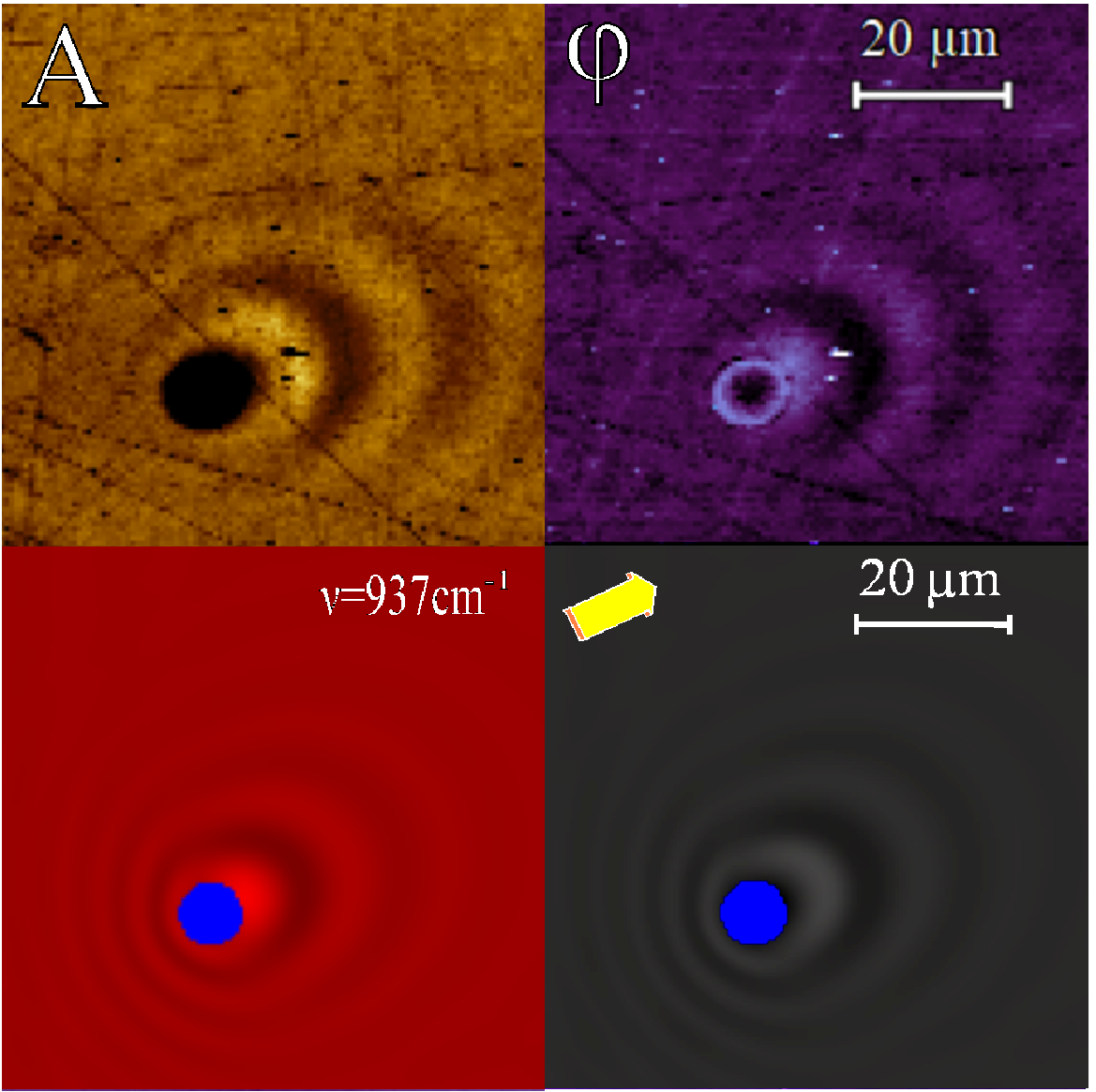}}
 \caption{\label{fig:RoundIsland}
Amplitude (left panels) and phase (right panels) of sSNOM signal
collected (top panels)  with an external irradiation near a round
island of an $Au$ mask (diameter 8 $\mu$m) on the $SiC$ surface.
Bottom panels show the Green-function based-simulations. Laser beam
direction is shown with an arrow. }
\end{figure}

 Importantly, the SPhP waves are generated not at the metal edges,
but rather over the whole metal-uncovered
area~\cite{LST_1941,Barron_PR1961,Rup_Englm_RPP1970,Mills_Burst_RPP1974}.
Without the metal mask, these elementary circular waves are emitted
in all directions almost isotropically, having a small total
momentum. Metal-covered areas introduce asymmetry to the problem and
make the momentum uncompensated.

To demonstrate the usefulness of the presented model, let us
consider an example of the wave pattern formed by a small $Au$
island measured simalarly to
Ref.~\cite{Hillenbr_Round_Island_SiC_JM2008}, see
fig.\ref{fig:RoundIsland}. In that paper, in order to explain the
observations, the oscillating dipoles excited by pumping in the gold
island and then emitting SPhP waves along the surface were
artificially introduced. The SNOM signal is the sum of a plane wave
incident directly from the laser and polariton waves propagating
along the surface. Therefore, at some points, where the phase of the
wave concentrically diverging from the island and the laser wave are
opposite, their sum should be equal to zero. The geometric location
of such points should be a plane hyperbola. However, hyperbolic
"calm" region is experimentally not observed. Another disadvantage
of the dipole model is the absence of an explanation for why the
amplitude of the wave launched by the dipole backwards the pump
radiation is significantly smaller than the amplitude of the wave
excited in the direction of the laser beam.

Calculations with the Green function straightforwardly resolve the
above contradictions and allow simulations of the SPhP wave patterns
excited on the $SiC$ surface by resonant light in the presence of a
metal mask without additional assumptions. The SPhP wave field is
straightforwardly obtained by integration (Eq. \ref{eq:GF_Integral})
for the known diameter of the island. The result of such a
simulation is shown in Fig.\ref{fig:RoundIsland} and agrees well
with the experimental data and, of course, has no hyperbolic "calm"
region.

Reflections of waves from the boundaries of the golden masks at the
edge of the window (black rays of polariton waves in
Fig.\ref{fig:sSNOM_Sample_n_Tip}) and their further propagation over
the $SiC$ surface in the mask window make the calculations
insurmountably complicated.

\begin{figure}
\resizebox{0.45\textwidth}{!}
 {\includegraphics{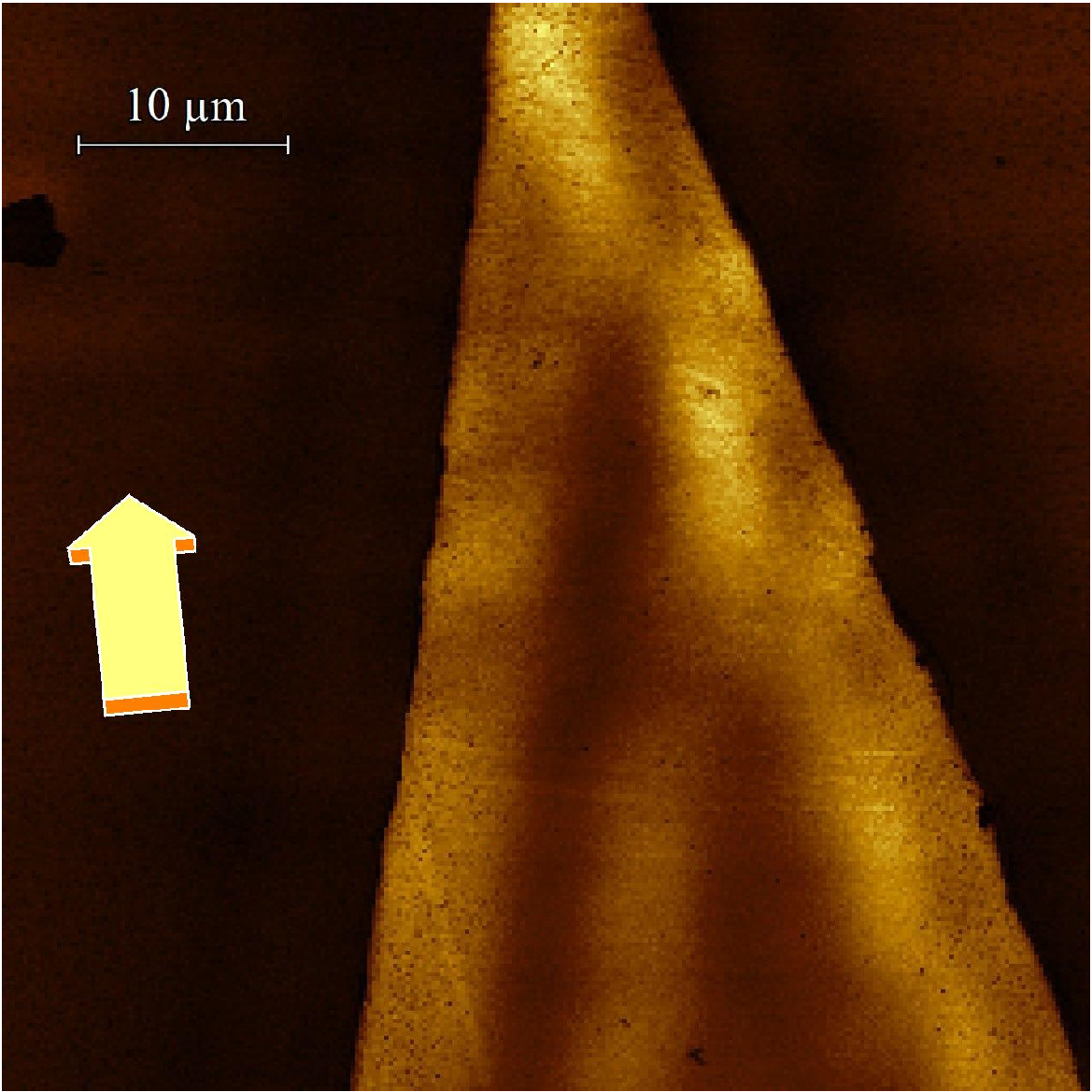}}
 \caption{\label{fig:V_corner}
Image of the amplitude of the sSNOM signal obtained on an open,
acute-angled region of the $SiC$ surface, on which the incident (in
the direction shown by the arrow) light wave excites running
phonon-polariton waves. The $CO_2$ laser line used ($\nu=935$
cm$^{-1}$). Waves with phase fronts parallel to the source edge are
visible. Waves with phase front directions corresponding to
geometric reflection are absent.}
\end{figure}

 {\bf Experimental result.}
 In this paper, we present an experimental observation {and
qualitative considerations} in favor of the absence of noticeable
reflections of the surface polariton wave from the edge of a metal
mask. The straight edge of the mask in the presence of a plane light
wave incident on the sample turns out to be a source of a polariton
wave traveling along the
surface~\cite{Sievers_SPP_Wave_Edge_Launch_APL_1978}, the wave front
of which is also straight. In sSNOM images of this traveling wave,
the phase front turns out to be parallel to the mask boundary
\cite{Huber_SiC_SPP_dispersion_APL2005, Kaza_JETP2006_Engl,
Kaza_Spiral_Trajectory_APA_2013}.

 If the SPhP wave were reflected from the straight edge of the metal
mask on the sample surface, a series of beating fringes
corresponding to the wave propagating in a new direction should
appear. We formed (Fig.\ref{fig:V_corner}) the following structure
on the surface of the $SiC$ sample: two straight edges of the metal
($Au$, 50~nm) mask, between which there is an acute corner of the
$SiC$ surface open to irradiation. In the experiment, two series of
beating fringes are observed between the laser beam field
$E_{las}(x,y)$ and the fields $E_{SPhP}^{(1)}(x,y)$ and
$E_{SPhP}^{(2)}(x,y)$ of polariton waves launched from the two
straight edges of the gold mask. The reflection of a wave, initially
launched by one edge, from the opposite edge of the triangular
~$SiC$ "bay" should have a third direction, not coinciding with
either of the two listed. No such wave is observed, indicating the
absence of reflection.

 {\bf Discussion.} Another, indirect experimental indication of the
absence of a noticeable reflection of the phonon-polariton wave
running along the $SiC$ surface from the boundary of the
polariton-active area is given by a comparison with systems in which
such a reflection does exist. For example, in sSNOM images of
nanometer-thick samples near the boundary of a film of
polariton-active substances
(graphene~\cite{Hillenbr_NatLett_Graphene_2012}, boron
nitride~\cite{Basov_Graphene_on_BN_NatNano_2015}, $MoS_2$
~\cite{SurfPhPolarit_WdW_NatComm_2018},
germanium~\cite{UltraConfined_PhPolarit_NatComm_2020}), there is a
series of parallel fringes with a period that weakly depends on the
orientation of the film edge with respect to the direction of the
exciting light. Such a wave pattern corresponds to a surface
polariton wave, which was launched by the  probing tip and then
reflected from the edge of the polariton-active film. After
reflection from the straight line of the boundary, this polariton
wave diverging from the "image" tip returned back to the probe and
modified the tip's electromagnetic response with its field, changing
the parameters of the light wave emitted by the probe into space.
The observed series of interference fringes of the probing tip field
with its own reflection at the edge of the film quickly fades away
with distance from the edge of the film, since the diverging
concentric wave is described by the cylindrical Hankel function (its
amplitude decreases approximately inversely proportional to the
distance from the source), and, thus, the amplitude of the field of
the wave received from the mirror image of the tip at the boundary
of the film should decrease according to the same law. A similar
series of rapidly decaying waves near the edge of the metal mask was
not observed in experiments on the $SiC$ surface
~\cite{Huber_SiC_SPP_dispersion_APL2005, Kaza_JETP2006_Engl,
Hillenbr_Horseshoe_APL2008, Hillenbr_Round_Island_SiC_JM2008,
Kaza_Bragg_APL_2024, Kaza_Spiral_Trajectory_APA_2013}.

It is important to highlight the difference between light-excited
plasmon-polariton in metal structures on dielectric substrate (for
example
\cite{Sievers_SPP_Wave_Edge_Launch_APL_1978,Double_Au_Disks_NL_2004,Muskens_DoubleDisks_OE_2007})
and SPhP wave on the metallized $SiC$. For plasmon polariton the
edge of the polariton-active medium (metal film) turns out to be
spatially sharp for the entire depth of the film. Therefore surface
wave reflection is effective and integration of the Green function
does not bring satisfactory description. This is opposite to the
SPhP case described in this paper. We would also expect high
reflection of the SPhP waves from the boundary of a 100~nm-thick
$SiC$ film mentioned in~\cite{Mancini_AcsPhot_2022}. Such a
thickness is significantly less than an SPhP wave penetration depth
in $SiC$, so that the electromagnetic boundary of the slab happens
at the same edge line for the whole depth.

 Let us consider theoretical arguments in favor of the absence of
noticeable reflection of the SPhP wave on the $SiC$ surface from the
edge of the metal mask. The depth of exponential decay of the field
in the $z$-direction, perpendicular to the surface, is estimated as
follows \cite{LST_1941,Barron_PR1961,
Rup_Englm_RPP1970,Mills_Burst_RPP1974}:
 \begin{equation}
 \delta_{z(bn)}(\omega)=\frac{\omega}{c}
 \frac{\varepsilon_{bn}(\omega)}{\sqrt{\varepsilon_{bn}(\omega)+\varepsilon_{ab}}}
 \label{eq:deltaZbelow}
 \end{equation}

When substituting into (\ref{eq:deltaZbelow}) the expression
(\ref{eq:EpsilonSiC}), this depth is, depending on the frequency,
from $0.06 \lambda=70nm$ at the lattice resonance frequency
$\nu_T=795cm^{-1}$ to $0.5 \lambda=6\mu m$ near the frequency of the
edge of the polariton state band, approximately equal to
$\nu_L=961cm^{-1}$ ~\cite{Eps_SiC_Landolt_B}. The appearance of a
sharp edge of the mask over the polariton wave along the surface
does not immediately have a strong influence on it, because the vast
majority of electromagnetic energy is concentrated in crystal at a
depth of up to $5 \mu m$. In three-dimensional (optics) and
one-dimensional (cable) wave processes, effective reflection of a
wave from the boundary of a medium occurs in cases where the spatial
scale of the step in permittivity, or wave resistance, is smaller
than the wavelength. In optics, such a case is described by the
Fresnel formulas. The reflection coefficient of a radio signal from
a broken or short-circuited end of a cable is also easily calculated
using the known impedance of the short circuit at a point. In the
case where the wave properties change "adiabatically smoothly" over
a segment comparable to or greater than the wavelength, the wave
reflection weakens or disappears completely.

 \begin{figure}
 \resizebox{0.45\textwidth}{!}
 {\includegraphics{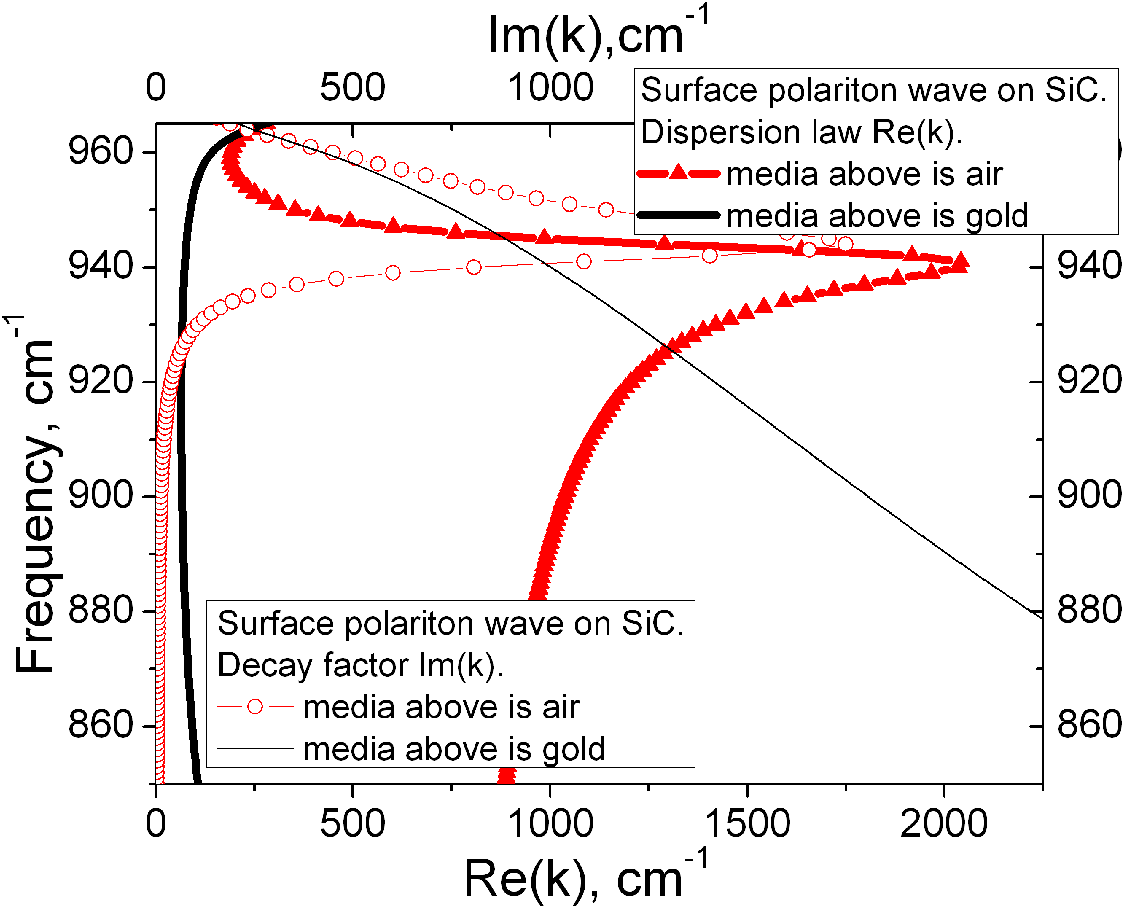}}
 \caption{ Dependence of the real (bold lines) and imaginary (thin
 lines) parts of the wave vector (horizontal scale) on the frequency
 (vertical scale) for phonon-polariton waves on the surface of $SiC$
 under vacuum (red curves) and under gold (black curves).
 }
 \label{fig:DispersionLaw}
\end{figure}

 The propagation length of the polariton wave under the gold mask can
be estimated by substituting  the upper medium
$\varepsilon_{ab}(\omega)$ material parameters for the gold film
\cite{Ordal_AO1985_Optical_properties_14-Metals} into the expression
(\ref{eq:k_xy}). This approach is consistent with the logic of
classical works on the theory of phonon-polariton
waves~\cite{LST_1941,Barron_PR1961, Rup_Englm_RPP1970,
Mills_Burst_RPP1974} and does not contradict any general
assumptions. Fig.\ref{fig:DispersionLaw} shows the result of the
calculations.

It can be seen in the figure \ref{fig:DispersionLaw}, that (a) the
phase velocity of the polariton wave under gold is very high (i.e.,
that the polariton oscillations in the "wave" that has penetrated
under the gold layer have the same phase everywhere), and (b) the
wave decay length is comparable to the wavelength over the open
surface of $SiC$.

{\bf Conclusion.} In this work, it has been experimentally shown
that the reflection of phonon-polariton waves running on the $SiC$
surface from the boundaries of the metal mask can be neglected. The
polariton waves, once excited by irradiating light on the surface,
disappear, getting under the edge of the mask. The first indirect
indications in favor of absence of reflections were repeatedly
reproducible agreements of the experimentally observed wavefields
with a Green function-based wavefront calculation for the metalized
masks of very complex shape on the SiC surface under resonant
irradiation~\cite{Huber_SiC_SPP_dispersion_APL2005,
Kaza_JETP2006_Engl,Kaza_Bragg_APL_2024}. In this paper we suggested
an experiment-crucious with a wedge-shaped mask and did not detect
polariton waves launched by one straight edge of the mask and then
reflected by the other one.
 This property of the polar crystal surface radically differs from
that of metal films used as carriers of plasmon-polariton waves in
the visible region of the spectrum, as well as from nanometer
graphene films and VdW crystals studied in the infrared region of
the spectrum. Due to this property, the configuration of standing
and running phonon-polariton waves excited by an incident external
coherent light wave in the presence of a metal screening mask of an
arbitrary configuration can be calculated by integrating the Green's
function.

\textbf{Acknowledgements} The sample fabrication was performed at
the Shared Facility center at the P.N. Lebedev Physical Institute.

\nocite{*}
\bibliography{Kaza_V_corner}

\providecommand{\noopsort}[1]{}\providecommand{\singleletter}[1]{#1}%
\begin{thebibliography}{31}
\expandafter\ifx\csname natexlab\endcsname\relax\def\natexlab#1{#1}\fi
\expandafter\ifx\csname bibnamefont\endcsname\relax
  \def\bibnamefont#1{#1}\fi
\expandafter\ifx\csname bibfnamefont\endcsname\relax
  \def\bibfnamefont#1{#1}\fi
\expandafter\ifx\csname citenamefont\endcsname\relax
  \def\citenamefont#1{#1}\fi
\expandafter\ifx\csname url\endcsname\relax
  \def\url#1{\texttt{#1}}\fi
\expandafter\ifx\csname urlprefix\endcsname\relax\def\urlprefix{URL }\fi
\providecommand{\bibinfo}[2]{#2}
\providecommand{\eprint}[2][]{\url{#2}}

\bibitem[{\citenamefont{{Basov} et~al.}(2021)\citenamefont{{Basov},
  {Asenjo-Garcia}, {Schuck}, {Zhu}, and
  {Rubio}}}]{Basov_PolaritonPanorama_NanoPhot_2021}
\bibinfo{author}{\bibfnamefont{D.~N.} \bibnamefont{{Basov}}},
  \bibinfo{author}{\bibfnamefont{A.}~\bibnamefont{{Asenjo-Garcia}}},
  \bibinfo{author}{\bibfnamefont{P.~J.} \bibnamefont{{Schuck}}},
  \bibinfo{author}{\bibfnamefont{X.}~\bibnamefont{{Zhu}}}, \bibnamefont{and}
  \bibinfo{author}{\bibfnamefont{A.}~\bibnamefont{{Rubio}}},
  \bibinfo{journal}{Nanophotonics} \textbf{\bibinfo{volume}{10}},
  \bibinfo{eid}{449} (\bibinfo{year}{2021}),
  \urlprefix\url{https://www.degruyter.com/document/doi/10.1515/nanoph-2020-0449/html}.

\bibitem[{\citenamefont{Sasaki et~al.}(1989)\citenamefont{Sasaki, Nishina,
  Sato, and Okamura}}]{Sasaki_SiC_Raman_PRB1989}
\bibinfo{author}{\bibfnamefont{Y.}~\bibnamefont{Sasaki}},
  \bibinfo{author}{\bibfnamefont{Y.}~\bibnamefont{Nishina}},
  \bibinfo{author}{\bibfnamefont{M.}~\bibnamefont{Sato}}, \bibnamefont{and}
  \bibinfo{author}{\bibfnamefont{K.}~\bibnamefont{Okamura}},
  \bibinfo{journal}{Phys. Rev. B} \textbf{\bibinfo{volume}{40}},
  \bibinfo{pages}{1762} (\bibinfo{year}{1989}),
  \urlprefix\url{http://link.aps.org/doi/10.1103/PhysRevB.40.1762}.

\bibitem[{\citenamefont{Bimberg et~al.}(1981)\citenamefont{Bimberg, Blachnik,
  Dean, Grave, Harbeke, H{\"u}bner, Kaufmann, Kress, Madelung
  et~al.}}]{Eps_SiC_Landolt_B}
\bibinfo{author}{\bibfnamefont{D.}~\bibnamefont{Bimberg}},
  \bibinfo{author}{\bibfnamefont{R.}~\bibnamefont{Blachnik}},
  \bibinfo{author}{\bibfnamefont{P.}~\bibnamefont{Dean}},
  \bibinfo{author}{\bibfnamefont{T.}~\bibnamefont{Grave}},
  \bibinfo{author}{\bibfnamefont{G.}~\bibnamefont{Harbeke}},
  \bibinfo{author}{\bibfnamefont{K.}~\bibnamefont{H{\"u}bner}},
  \bibinfo{author}{\bibfnamefont{U.}~\bibnamefont{Kaufmann}},
  \bibinfo{author}{\bibfnamefont{W.}~\bibnamefont{Kress}},
  \bibinfo{author}{\bibfnamefont{O.}~\bibnamefont{Madelung}},
  \bibnamefont{et~al.}, \emph{\bibinfo{title}{Physics of Group IV Elements and
  III-V Compounds / Physik Der Elemente Der IV. Gruppe und Der III-V
  Verbindungen}}, Landolt-Bornstein: Numerical Data and Functional Relationshi
  (\bibinfo{publisher}{Springer}, \bibinfo{year}{1981}), ISBN
  \bibinfo{isbn}{9783540106104},
  \urlprefix\url{http://books.google.de/books?id=6Ar774laEywC}.

\bibitem[{\citenamefont{Lyddane et~al.}(1941)\citenamefont{Lyddane, Sachs, and
  Teller}}]{LST_1941}
\bibinfo{author}{\bibfnamefont{R.~H.} \bibnamefont{Lyddane}},
  \bibinfo{author}{\bibfnamefont{R.~G.} \bibnamefont{Sachs}}, \bibnamefont{and}
  \bibinfo{author}{\bibfnamefont{E.}~\bibnamefont{Teller}},
  \bibinfo{journal}{Phys. Rev.} \textbf{\bibinfo{volume}{59}},
  \bibinfo{pages}{673} (\bibinfo{year}{1941}).

\bibitem[{\citenamefont{Barron}(1961)}]{Barron_PR1961}
\bibinfo{author}{\bibfnamefont{T.~H.~K.} \bibnamefont{Barron}},
  \bibinfo{journal}{Phys. Rev.} \textbf{\bibinfo{volume}{123}},
  \bibinfo{pages}{1995} (\bibinfo{year}{1961}),
  \urlprefix\url{http://link.aps.org/doi/10.1103/PhysRev.123.1995}.

\bibitem[{\citenamefont{Ruppin and Englman}(1970)}]{Rup_Englm_RPP1970}
\bibinfo{author}{\bibfnamefont{R.}~\bibnamefont{Ruppin}} \bibnamefont{and}
  \bibinfo{author}{\bibfnamefont{R.}~\bibnamefont{Englman}},
  \bibinfo{journal}{Rep. Prog. Phys} \textbf{\bibinfo{volume}{33}},
  \bibinfo{pages}{149} (\bibinfo{year}{1970}).

\bibitem[{\citenamefont{Mills and Burstein}(1974)}]{Mills_Burst_RPP1974}
\bibinfo{author}{\bibfnamefont{D.~L.} \bibnamefont{Mills}} \bibnamefont{and}
  \bibinfo{author}{\bibfnamefont{E.}~\bibnamefont{Burstein}},
  \bibinfo{journal}{Reports on Progress in Physics}
  \textbf{\bibinfo{volume}{37}}, \bibinfo{pages}{817} (\bibinfo{year}{1974}),
  \urlprefix\url{http://iopscience.iop.org/0034-4885/37/7/001/}.

\bibitem[{\citenamefont{Huber et~al.}(2005)\citenamefont{Huber, Ocelic,
  Kazantsev, and Hillenbrand}}]{Huber_SiC_SPP_dispersion_APL2005}
\bibinfo{author}{\bibfnamefont{A.}~\bibnamefont{Huber}},
  \bibinfo{author}{\bibfnamefont{N.}~\bibnamefont{Ocelic}},
  \bibinfo{author}{\bibfnamefont{D.}~\bibnamefont{Kazantsev}},
  \bibnamefont{and}
  \bibinfo{author}{\bibfnamefont{R.}~\bibnamefont{Hillenbrand}},
  \bibinfo{journal}{Applied Physics Letters} \textbf{\bibinfo{volume}{87}},
  \bibinfo{eid}{081103} (pages~\bibinfo{numpages}{3}) (\bibinfo{year}{2005}),
  \urlprefix\url{https://pubs.aip.org/aip/apl/article/87/8/081103/117557/Near-field-imaging-of-mid-infrared-surface-phonon}.

\bibitem[{\citenamefont{Mancini et~al.}(2022)\citenamefont{Mancini, Nan,
  Wendisch, Bert\'e, Ren, Cort\'es, and Maier}}]{Mancini_AcsPhot_2022}
\bibinfo{author}{\bibfnamefont{A.}~\bibnamefont{Mancini}},
  \bibinfo{author}{\bibfnamefont{L.}~\bibnamefont{Nan}},
  \bibinfo{author}{\bibfnamefont{F.}~\bibnamefont{Wendisch}},
  \bibinfo{author}{\bibfnamefont{R.}~\bibnamefont{Bert\'e}},
  \bibinfo{author}{\bibfnamefont{H.}~\bibnamefont{Ren}},
  \bibinfo{author}{\bibfnamefont{E.}~\bibnamefont{Cort\'es}}, \bibnamefont{and}
  \bibinfo{author}{\bibfnamefont{S.}~\bibnamefont{Maier}},
  \bibinfo{journal}{ACS Photonics} \textbf{\bibinfo{volume}{9}},
  \bibinfo{pages}{3696} (\bibinfo{year}{2022}),
  \urlprefix\url{https://doi.org/10.1021/acsphotonics.2c01270}.

\bibitem[{\citenamefont{Dubrovkin et~al.}(2020)\citenamefont{Dubrovkin, Qiang,
  Salim, Nam, Zheludev, and Wang}}]{UltraConfined_PhPolarit_NatComm_2020}
\bibinfo{author}{\bibfnamefont{A.~M.} \bibnamefont{Dubrovkin}},
  \bibinfo{author}{\bibfnamefont{B.}~\bibnamefont{Qiang}},
  \bibinfo{author}{\bibfnamefont{T.}~\bibnamefont{Salim}},
  \bibinfo{author}{\bibfnamefont{D.}~\bibnamefont{Nam}},
  \bibinfo{author}{\bibfnamefont{N.~I.} \bibnamefont{Zheludev}},
  \bibnamefont{and} \bibinfo{author}{\bibfnamefont{Q.~J.} \bibnamefont{Wang}},
  \bibinfo{journal}{Nature Communications} \textbf{\bibinfo{volume}{11}},
  \bibinfo{pages}{1863} (\bibinfo{year}{2020}), ISSN \bibinfo{issn}{2041-1723},
  \urlprefix\url{https://doi.org/10.1038/s41467-020-15767-y}.

\bibitem[{\citenamefont{Huber et~al.}(2008{\natexlab{a}})\citenamefont{Huber,
  Ocelic, and Hillenbrand}}]{Hillenbr_Round_Island_SiC_JM2008}
\bibinfo{author}{\bibfnamefont{A.}~\bibnamefont{Huber}},
  \bibinfo{author}{\bibfnamefont{N.}~\bibnamefont{Ocelic}}, \bibnamefont{and}
  \bibinfo{author}{\bibfnamefont{R.}~\bibnamefont{Hillenbrand}},
  \bibinfo{journal}{Journal of Microscopy} \textbf{\bibinfo{volume}{229}},
  \bibinfo{pages}{389} (\bibinfo{year}{2008}{\natexlab{a}}),
  \urlprefix\url{http://dx.doi.org/10.1111/j.1365-2818.2008.01917.x}.

\bibitem[{\citenamefont{Huber et~al.}(2008{\natexlab{b}})\citenamefont{Huber,
  Deutsch, Novotny, and Hillenbrand}}]{Hillenbr_Horseshoe_APL2008}
\bibinfo{author}{\bibfnamefont{A.~J.} \bibnamefont{Huber}},
  \bibinfo{author}{\bibfnamefont{B.}~\bibnamefont{Deutsch}},
  \bibinfo{author}{\bibfnamefont{L.}~\bibnamefont{Novotny}}, \bibnamefont{and}
  \bibinfo{author}{\bibfnamefont{R.}~\bibnamefont{Hillenbrand}},
  \bibinfo{journal}{Applied Physics Letters} \textbf{\bibinfo{volume}{92}},
  \bibinfo{eid}{203104} (pages~\bibinfo{numpages}{3})
  (\bibinfo{year}{2008}{\natexlab{b}}),
  \urlprefix\url{http://link.aip.org/link/?APL/92/203104/1}.

\bibitem[{\citenamefont{Ivchenko et~al.}(2024)\citenamefont{Ivchenko,
  Kazantsev, Ievleva, Kazantseva, and Kuntsevich}}]{Kaza_Bragg_APL_2024}
\bibinfo{author}{\bibfnamefont{V.~S.} \bibnamefont{Ivchenko}},
  \bibinfo{author}{\bibfnamefont{D.~V.} \bibnamefont{Kazantsev}},
  \bibinfo{author}{\bibfnamefont{V.~A.} \bibnamefont{Ievleva}},
  \bibinfo{author}{\bibfnamefont{E.~A.} \bibnamefont{Kazantseva}},
  \bibnamefont{and} \bibinfo{author}{\bibfnamefont{A.~Y.}
  \bibnamefont{Kuntsevich}}, \bibinfo{journal}{Applied Physics Letters}
  \textbf{\bibinfo{volume}{125}}, \bibinfo{pages}{171601}
  (\bibinfo{year}{2024}), ISSN \bibinfo{issn}{0003-6951},
  \eprint{https://pubs.aip.org/aip/apl/article-pdf/doi/10.1063/5.0229574/20219776/171601\_1\_5.0229574.pdf},
  \urlprefix\url{https://doi.org/10.1063/5.0229574}.

\bibitem[{\citenamefont{Chen et~al.}(2012)\citenamefont{Chen, Badioli,
  Alonso-Gonzalez, Thongrattanasiri, Centeno, Pesquera, Godignon, Elorza,
  Camara, de~Abajo et~al.}}]{Hillenbr_NatLett_Graphene_2012}
\bibinfo{author}{\bibfnamefont{J.}~\bibnamefont{Chen}},
  \bibinfo{author}{\bibfnamefont{M.}~\bibnamefont{Badioli}},
  \bibinfo{author}{\bibfnamefont{P.}~\bibnamefont{Alonso-Gonzalez}},
  \bibinfo{author}{\bibfnamefont{S.}~\bibnamefont{Thongrattanasiri}},
  \bibinfo{author}{\bibfnamefont{F.~H. J. O. M. S.~A.} \bibnamefont{Centeno}},
  \bibinfo{author}{\bibfnamefont{A.}~\bibnamefont{Pesquera}},
  \bibinfo{author}{\bibfnamefont{P.}~\bibnamefont{Godignon}},
  \bibinfo{author}{\bibfnamefont{A.~Z.} \bibnamefont{Elorza}},
  \bibinfo{author}{\bibfnamefont{N.}~\bibnamefont{Camara}},
  \bibinfo{author}{\bibfnamefont{F.~J.~G.} \bibnamefont{de~Abajo}},
  \bibnamefont{et~al.}, \bibinfo{journal}{Nature Letters}
  \textbf{\bibinfo{volume}{487}}, \bibinfo{pages}{77}
  (\bibinfo{year}{2012}),
  \urlprefix\url{http://dx.doi.org/10.1038/nature11254}.

\bibitem[{\citenamefont{Dai et~al.}(2015)\citenamefont{Dai, Ma, Liu, Andersen,
  Fei, Goldflam, Wagner, Watanabe, Taniguchi, Thiemens
  et~al.}}]{Basov_Graphene_on_BN_NatNano_2015}
\bibinfo{author}{\bibfnamefont{S.}~\bibnamefont{Dai}},
  \bibinfo{author}{\bibfnamefont{Q.}~\bibnamefont{Ma}},
  \bibinfo{author}{\bibfnamefont{M.~K.} \bibnamefont{Liu}},
  \bibinfo{author}{\bibfnamefont{T.}~\bibnamefont{Andersen}},
  \bibinfo{author}{\bibfnamefont{Z.}~\bibnamefont{Fei}},
  \bibinfo{author}{\bibfnamefont{M.~D.} \bibnamefont{Goldflam}},
  \bibinfo{author}{\bibfnamefont{M.}~\bibnamefont{Wagner}},
  \bibinfo{author}{\bibfnamefont{K.}~\bibnamefont{Watanabe}},
  \bibinfo{author}{\bibfnamefont{T.}~\bibnamefont{Taniguchi}},
  \bibinfo{author}{\bibfnamefont{M.}~\bibnamefont{Thiemens}},
  \bibnamefont{et~al.}, \bibinfo{journal}{Nature Nanotechnology}
  \textbf{\bibinfo{volume}{10}}, \bibinfo{pages}{682} (\bibinfo{year}{2015}),
  ISSN \bibinfo{issn}{1748-3395},
  \urlprefix\url{https://doi.org/10.1038/nnano.2015.131}.

\bibitem[{\citenamefont{Dubrovkin et~al.}(2018)\citenamefont{Dubrovkin, Qiang,
  Krishnamoorthy, Zheludev, and Wang}}]{SurfPhPolarit_WdW_NatComm_2018}
\bibinfo{author}{\bibfnamefont{A.~M.} \bibnamefont{Dubrovkin}},
  \bibinfo{author}{\bibfnamefont{B.}~\bibnamefont{Qiang}},
  \bibinfo{author}{\bibfnamefont{H.~N.~S.} \bibnamefont{Krishnamoorthy}},
  \bibinfo{author}{\bibfnamefont{N.~I.} \bibnamefont{Zheludev}},
  \bibnamefont{and} \bibinfo{author}{\bibfnamefont{Q.~J.} \bibnamefont{Wang}},
  \bibinfo{journal}{Nature Communications} \textbf{\bibinfo{volume}{9}},
  \bibinfo{pages}{1762} (\bibinfo{year}{2018}), ISSN \bibinfo{issn}{2041-1723},
  \urlprefix\url{https://doi.org/10.1038/s41467-018-04168-x}.

\bibitem[{\citenamefont{Wickramasinghe and
  Williams}(1990)}]{ASNOM_Wickramasinghe_PatentUSA_1990}
\bibinfo{author}{\bibfnamefont{H.}~\bibnamefont{Wickramasinghe}}
  \bibnamefont{and} \bibinfo{author}{\bibfnamefont{C.}~\bibnamefont{Williams}},
  \emph{\bibinfo{title}{Apertureless near field optical microscope}}
  (\bibinfo{year}{1990}), \bibinfo{note}{uS Patent 4,947,034},
  \urlprefix\url{http://www.google.com/patents/US4947034}.

\bibitem[{\citenamefont{Zenhausern et~al.}(1994)\citenamefont{Zenhausern,
  O'Boyle, and Wickramasinghe}}]{s_SNOM_first}
\bibinfo{author}{\bibfnamefont{F.}~\bibnamefont{Zenhausern}},
  \bibinfo{author}{\bibfnamefont{M.~P.} \bibnamefont{O'Boyle}},
  \bibnamefont{and} \bibinfo{author}{\bibfnamefont{H.~K.}
  \bibnamefont{Wickramasinghe}}, \bibinfo{journal}{Applied Physics Letters}
  \textbf{\bibinfo{volume}{65}}, \bibinfo{pages}{1623} (\bibinfo{year}{1994}),
  \urlprefix\url{http://link.aip.org/link/?APL/65/1623/1}.

\bibitem[{\citenamefont{Labardi et~al.}(2000)\citenamefont{Labardi, Patan\`{e},
  and Allegrini}}]{Labardi_SecondHarm_ASNOM_APL2000}
\bibinfo{author}{\bibfnamefont{M.}~\bibnamefont{Labardi}},
  \bibinfo{author}{\bibfnamefont{S.}~\bibnamefont{Patan\`{e}}},
  \bibnamefont{and}
  \bibinfo{author}{\bibfnamefont{M.}~\bibnamefont{Allegrini}},
  \bibinfo{journal}{Applied Physics Letters} \textbf{\bibinfo{volume}{77}},
  \bibinfo{pages}{621} (\bibinfo{year}{2000}),
  \urlprefix\url{http://link.aip.org/link/?APL/77/621/1}.

\bibitem[{\citenamefont{Keilmann and
  Hillenbrand}(2004)}]{Keilmann_PTRS2004_sSNOM}
\bibinfo{author}{\bibfnamefont{F.}~\bibnamefont{Keilmann}} \bibnamefont{and}
  \bibinfo{author}{\bibfnamefont{R.}~\bibnamefont{Hillenbrand}},
  \bibinfo{journal}{Philosophical Transactions: Mathematical, Physical and
  Engineering Sciences} \textbf{\bibinfo{volume}{362}}, \bibinfo{pages}{787}
  (\bibinfo{year}{2004}), ISSN \bibinfo{issn}{1364503X},
  \urlprefix\url{http://www.jstor.org/stable/4142390}.

\bibitem[{\citenamefont{Kazantsev}(2006)}]{Kaza_JETP2006_Engl}
\bibinfo{author}{\bibfnamefont{D.~V.} \bibnamefont{Kazantsev}},
  \bibinfo{journal}{JETP Letters} \textbf{\bibinfo{volume}{83}},
  \bibinfo{pages}{323} (\bibinfo{year}{2006}),
  \urlprefix\url{http://link.springer.com/article/10.1134%2FS0021364006080054}.

\bibitem[{\citenamefont{Kazantsev and
  Ryssel}(2013)}]{Kaza_Spiral_Trajectory_APA_2013}
\bibinfo{author}{\bibfnamefont{D.}~\bibnamefont{Kazantsev}} \bibnamefont{and}
  \bibinfo{author}{\bibfnamefont{H.}~\bibnamefont{Ryssel}},
  \bibinfo{journal}{Applied Physics A} \textbf{\bibinfo{volume}{113}},
  \bibinfo{pages}{27} (\bibinfo{year}{2013}),
  \urlprefix\url{http://dx.doi.org/10.1007/s00339-013-7869-y}.

\bibitem[{\citenamefont{Kazantsev and
  Kazantseva}(2018)}]{Kaza_FieldEnhancement_JETPLe_2018}
\bibinfo{author}{\bibfnamefont{D.~V.} \bibnamefont{Kazantsev}}
  \bibnamefont{and} \bibinfo{author}{\bibfnamefont{E.~A.}
  \bibnamefont{Kazantseva}}, \bibinfo{journal}{JETP Letters}
  \textbf{\bibinfo{volume}{107}}, \bibinfo{pages}{512} (\bibinfo{year}{2018}),
  ISSN \bibinfo{issn}{1090-6487},
  \urlprefix\url{https://doi.org/10.1134/S0021364018080106}.

\bibitem[{\citenamefont{Zayats et~al.}()\citenamefont{Zayats, Smolyaninov, and
  Maradudin}}]{Zayats_Plasmonics_PhysRep_2005}
\bibinfo{author}{\bibfnamefont{A.~V.} \bibnamefont{Zayats}},
  \bibinfo{author}{\bibfnamefont{I.~I.} \bibnamefont{Smolyaninov}},
  \bibnamefont{and} \bibinfo{author}{\bibfnamefont{A.~A.}
  \bibnamefont{Maradudin}}, \bibinfo{journal}{Physics Reports}
  \textbf{\bibinfo{volume}{408}}, \bibinfo{pages}{131} (????),
  \urlprefix\url{https://www.sciencedirect.com/science/article/abs/pii/S0370157304004600}.

\bibitem[{\citenamefont{Krenn et~al.}(1999)\citenamefont{Krenn, Dereux, Weeber,
  Bourillot, Lacroute, Goudonnet, Schider, Gotschy, Leitner, Aussenegg
  et~al.}}]{Krenn_PlasmonGF_PRL_1999}
\bibinfo{author}{\bibfnamefont{J.~R.} \bibnamefont{Krenn}},
  \bibinfo{author}{\bibfnamefont{A.}~\bibnamefont{Dereux}},
  \bibinfo{author}{\bibfnamefont{J.~C.} \bibnamefont{Weeber}},
  \bibinfo{author}{\bibfnamefont{E.}~\bibnamefont{Bourillot}},
  \bibinfo{author}{\bibfnamefont{Y.}~\bibnamefont{Lacroute}},
  \bibinfo{author}{\bibfnamefont{J.~P.} \bibnamefont{Goudonnet}},
  \bibinfo{author}{\bibfnamefont{G.}~\bibnamefont{Schider}},
  \bibinfo{author}{\bibfnamefont{W.}~\bibnamefont{Gotschy}},
  \bibinfo{author}{\bibfnamefont{A.}~\bibnamefont{Leitner}},
  \bibinfo{author}{\bibfnamefont{F.~R.} \bibnamefont{Aussenegg}},
  \bibnamefont{et~al.}, \bibinfo{journal}{Phys. Rev. Lett.}
  \textbf{\bibinfo{volume}{82}}, \bibinfo{pages}{2590} (\bibinfo{year}{1999}),
  \urlprefix\url{https://link.aps.org/doi/10.1103/PhysRevLett.82.2590}.

\bibitem[{\citenamefont{Feldman et~al.}(1968)\citenamefont{Feldman, Parker,
  Choyke, and Patrick}}]{RamanSiC_PR_1968}
\bibinfo{author}{\bibfnamefont{D.~W.} \bibnamefont{Feldman}},
  \bibinfo{author}{\bibfnamefont{J.~H.} \bibnamefont{Parker}},
  \bibinfo{author}{\bibfnamefont{W.~J.} \bibnamefont{Choyke}},
  \bibnamefont{and} \bibinfo{author}{\bibfnamefont{L.}~\bibnamefont{Patrick}},
  \bibinfo{journal}{Phys. Rev.} \textbf{\bibinfo{volume}{170}},
  \bibinfo{pages}{698} (\bibinfo{year}{1968}),
  \urlprefix\url{https://link.aps.org/doi/10.1103/PhysRev.170.698}.

\bibitem[{\citenamefont{Harima et~al.}(1995)\citenamefont{Harima, Nakashima,
  and Uemura}}]{Harima_AP-1995_Raman}
\bibinfo{author}{\bibfnamefont{H.}~\bibnamefont{Harima}},
  \bibinfo{author}{\bibfnamefont{S.}~\bibnamefont{Nakashima}},
  \bibnamefont{and} \bibinfo{author}{\bibfnamefont{T.}~\bibnamefont{Uemura}},
  \bibinfo{journal}{Journal of Applied Physics} \textbf{\bibinfo{volume}{78}},
  \bibinfo{pages}{1996} (\bibinfo{year}{1995}),
  \urlprefix\url{http://link.aip.org/link/?JAP/78/1996/1}.

\bibitem[{\citenamefont{Chabal and
  Sievers}(1978)}]{Sievers_SPP_Wave_Edge_Launch_APL_1978}
\bibinfo{author}{\bibfnamefont{Y.~J.} \bibnamefont{Chabal}} \bibnamefont{and}
  \bibinfo{author}{\bibfnamefont{A.~J.} \bibnamefont{Sievers}},
  \bibinfo{journal}{Applied Physics Letters} \textbf{\bibinfo{volume}{32}},
  \bibinfo{pages}{90} (\bibinfo{year}{1978}), ISSN \bibinfo{issn}{0003-6951},
  \eprint{https://pubs.aip.org/aip/apl/article-pdf/32/2/90/7739702/90\_1\_online.pdf},
  \urlprefix\url{https://doi.org/10.1063/1.89947}.

\bibitem[{\citenamefont{Atay et~al.}(2004)\citenamefont{Atay, Song, and
  Nurmikko}}]{Double_Au_Disks_NL_2004}
\bibinfo{author}{\bibfnamefont{T.}~\bibnamefont{Atay}},
  \bibinfo{author}{\bibfnamefont{J.-H.} \bibnamefont{Song}}, \bibnamefont{and}
  \bibinfo{author}{\bibfnamefont{A.~V.} \bibnamefont{Nurmikko}},
  \bibinfo{journal}{Nano Lett.} \textbf{\bibinfo{volume}{4}},
  \bibinfo{pages}{1627} (\bibinfo{year}{2004}), ISSN \bibinfo{issn}{1530-6984},
  \urlprefix\url{https://doi.org/10.1021/nl049215n}.

\bibitem[{\citenamefont{Muskens et~al.}(2007)\citenamefont{Muskens, Giannini,
  S\'{a}nchez-Gil, and Rivas}}]{Muskens_DoubleDisks_OE_2007}
\bibinfo{author}{\bibfnamefont{O.~L.} \bibnamefont{Muskens}},
  \bibinfo{author}{\bibfnamefont{V.}~\bibnamefont{Giannini}},
  \bibinfo{author}{\bibfnamefont{J.~A.} \bibnamefont{S\'{a}nchez-Gil}},
  \bibnamefont{and} \bibinfo{author}{\bibfnamefont{J.~G.} \bibnamefont{Rivas}},
  \bibinfo{journal}{Opt. Express} \textbf{\bibinfo{volume}{15}},
  \bibinfo{pages}{17736} (\bibinfo{year}{2007}),
  \urlprefix\url{https://opg.optica.org/oe/abstract.cfm?URI=oe-15-26-17736}.

\bibitem[{\citenamefont{Ordal et~al.}(1985)\citenamefont{Ordal, Bell,
  Alexander, Long, and Querry}}]{Ordal_AO1985_Optical_properties_14-Metals}
\bibinfo{author}{\bibfnamefont{M.~A.} \bibnamefont{Ordal}},
  \bibinfo{author}{\bibfnamefont{R.~J.} \bibnamefont{Bell}},
  \bibinfo{author}{\bibfnamefont{R.~W.} \bibnamefont{Alexander}},
  \bibinfo{author}{\bibfnamefont{L.~L.} \bibnamefont{Long}}, \bibnamefont{and}
  \bibinfo{author}{\bibfnamefont{M.~R.} \bibnamefont{Querry}},
  \bibinfo{journal}{Appl. Opt.} \textbf{\bibinfo{volume}{24}},
  \bibinfo{pages}{4493} (\bibinfo{year}{1985}),
  \urlprefix\url{http://ao.osa.org/abstract.cfm?URI=ao-24-24-4493}.

\end{thebibliography}
\end{document}